\documentclass[sigconf]{acmart}
\usepackage{booktabs} 
\usepackage{geometry}
\usepackage{multirow}
\usepackage{array, makecell, rotating, boldline}
\usepackage{dblfloatfix}
\usepackage{caption}
\usepackage{color}
\newcommand\nocell[1]{\multicolumn{#1}{c|}{}}
\setcellgapes{3pt}

\makeatletter
\def\blfootnote{\gdef\@thefnmark{}\@footnotetext}
\makeatother

\setcopyright{none}
\renewcommand\footnotetextcopyrightpermission[1]{}

\usepackage{hyperref}

\begin{document}
\title{Multimedia Semantic Integrity Assessment Using Joint Embedding Of Images And Text}


\author{Ayush Jaiswal\textsuperscript{$\ast$}}
\affiliation{
  \institution{USC Information Sciences Institute}
	\streetaddress{4676 Admiralty Way}
  \city{Marina del Rey} 
  \state{CA}
  \country{USA}
  \postcode{90292}
}
\email{ajaiswal@isi.edu}

\author{Ekraam Sabir\textsuperscript{$\ast$}}
\affiliation{
  \institution{USC Information Sciences Institute}
	\streetaddress{4676 Admiralty Way}
  \city{Marina del Rey}
  \state{CA}
  \country{USA}
  \postcode{90292}
}
\email{esabir@isi.edu}

\author{Wael AbdAlmageed}
\affiliation{
  \institution{USC Information Sciences Institute}
	\streetaddress{4676 Admiralty Way}
  \city{Marina del Rey} 
  \state{CA}
  \country{USA}
  \postcode{90292}
}
\email{wamageed@isi.edu}

\author{Premkumar Natarajan}
\affiliation{
  \institution{USC Information Sciences Institute}
	\streetaddress{4676 Admiralty Way}
  \city{Marina del Rey} 
  \state{CA}
  \country{USA}
  \postcode{90292}
}
\email{pnataraj@isi.edu}

\renewcommand{\shortauthors}{A. Jaiswal et al.}

\begin{abstract}
\blfootnote{\textsuperscript{$\ast$}Ayush Jaiswal and Ekraam Sabir contributed equally to the work in this paper.}

Real-world multimedia data is often composed of multiple modalities such as an image or a video with associated text (e.g., captions, user comments, etc.) and metadata. Such multimodal data \emph{packages} are prone to manipulations, where a subset of these modalities can be altered to misrepresent or repurpose data packages, with possible malicious intent. It is therefore important to develop methods to assess or verify the integrity of these multimedia packages. Using computer vision and natural language processing methods to directly compare the image (or video) and the associated caption to verify the integrity of a media package is only possible for a limited set of objects and scenes. In this paper we present a novel deep-learning-based approach that uses a reference set of multimedia packages to assess the semantic integrity of multimedia packages containing images and captions. We construct a joint embedding of images and captions with deep multimodal representation learning on the reference dataset in a framework that also provides image-caption consistency scores (ICCSs). The integrity of query media packages is assessed as the \textit{inlierness} of the query ICCSs with respect to the reference dataset. We present the MultimodAl Information Manipulation dataset (MAIM), a new dataset of media packages from Flickr, which we are making available to the research community. We use both the newly created dataset as well as Flickr30K and MS COCO datasets to quantitatively evaluate our proposed approach. The reference dataset does not contain \emph{unmanipulated} versions of tampered query packages. Our method is able to achieve $F_1$ scores of $0.75$, $0.89$ and $0.94$ on MAIM, Flickr30K and MS COCO, respectively, for detecting semantically incoherent media packages. 
\end{abstract}

\ccsdesc[300]{Computing methodologies~image representations}
\ccsdesc[300]{Computing methodologies~computer vision}
\ccsdesc[300]{Computing methodologies~natural language processing}
\ccsdesc[300]{Computing methodologies~unsupervised learning}
\ccsdesc[300]{Computing methodologies~neural networks}

\keywords{semantic integrity assemessment; multimedia data; multimodal semantic integrity; fake news}

\maketitle

\section{Introduction}

\begin{figure}[t]
\centering
\includegraphics[width=0.45\textwidth]{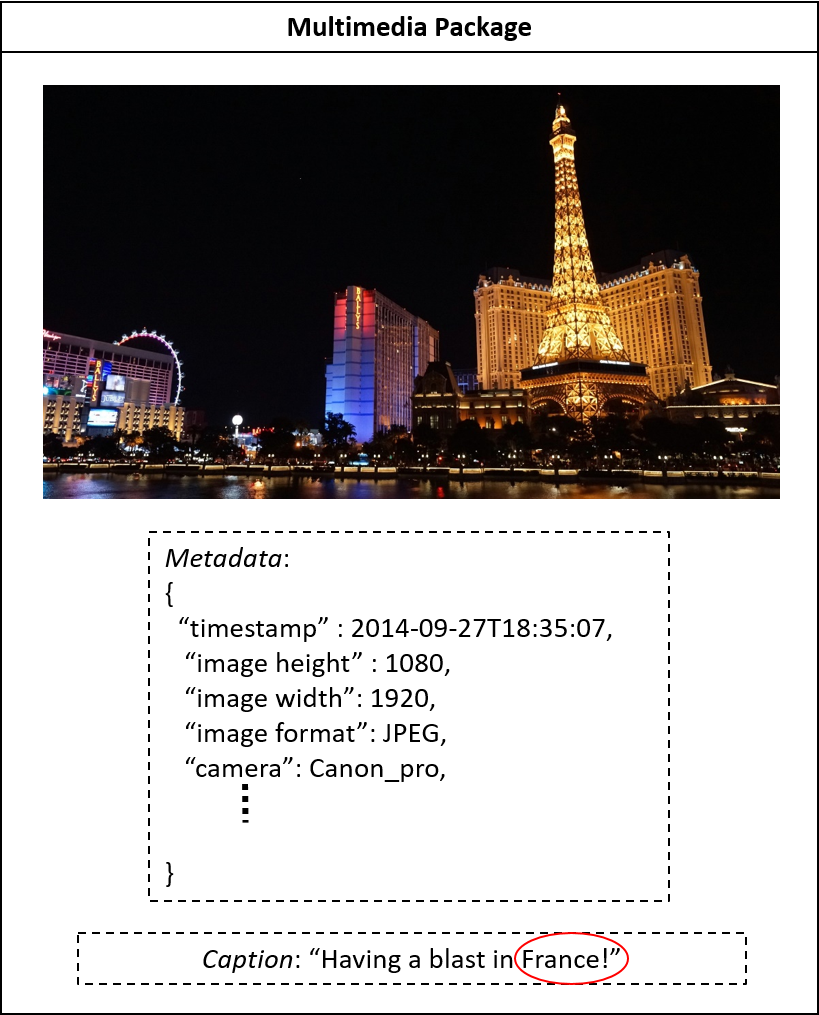}
\caption{Multimodal information manipulation example. This is a photograph of the Eiffel Tower in Las Vegas, but the caption says France.}
\label{fig:larger_problem_example}
\end{figure}

\begin{figure*}[t]
\centering
\includegraphics[width=0.95\textwidth]{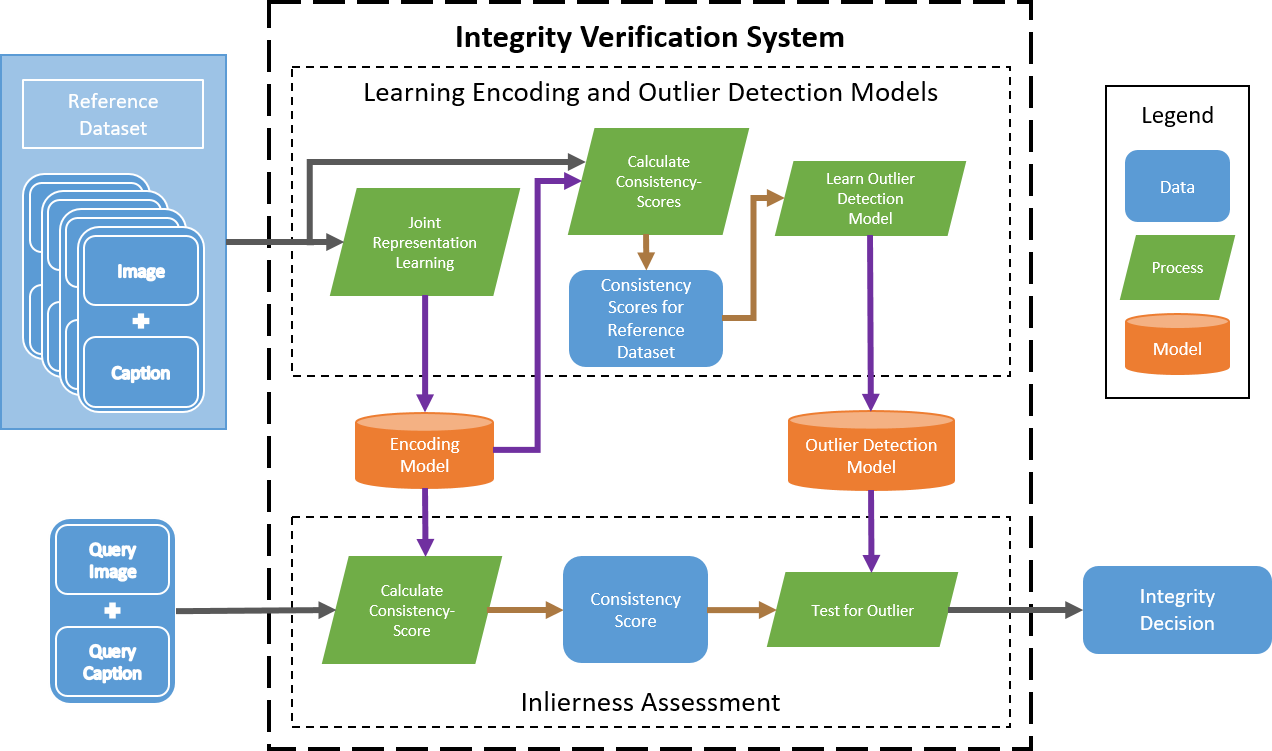}
\caption{Package Integrity Assessment System}
\label{fig:system}
\end{figure*}

In real life, data often presents itself with multiple modalities, where information about an entity or an event is incompletely captured by each modality separately. For example, a caption associated with the image of a person might provide information such as the name of the person and the location where the picture was taken, while other metadata might provide the date and time at which the image was taken. Independent existence of each modality makes multimedia data packages vulnerable to tampering, where the data in a subset of modalities of a multimedia package can be modified to misrepresent or repurpose the multimedia package. Such tampering, with possible malicious intent, can be misleading, if not dangerous. The location information, for example, in the aforementioned caption could be modified without an easy way to detect such tampering. Fig.~\ref{fig:larger_problem_example} demonstrates an example of information manipulation where a photograph of the Paris casino in Las Vegas, Nevada (which includes a half-scale replica of the Eiffel Tower)  is repurposed as a photograph of the real Eiffel Tower in Paris, France. Nevertheless, if the image has visual cues, such as a landmark, a person \textit{familiar} with the location can easily detect such a manipulation. However, this is a challenging multimedia analysis task, especially with the subtlety of data manipulation, the absence of clear visual cues (e.g., Eiffel Tower) and the proliferation of multimedia content from mobile devices and digital cameras. 

Verification of the integrity of information contained in any kind of data requires the existence of some form of prior knowledge. In the previous example, this knowledge is represented by a person's \textit{familiarity} with the location. Human beings use their knowledge, learned over time, or external sources such as encyclopedias, as knowledge bases (KBs). Motivated by this important observation, multimedia analysis algorithms could also take advantage of a KB to automatically assess the integrity of multimedia packages. A KB can be either implicit (such as a trained scene understanding and/or recognition model) or explicit (such as a database of known facts). In this paper we explore the use of a reference dataset (RD) of multimedia packages to assess the integrity of query packages. The RD is assumed to \emph{not} include other copies of the query image. Otherwise, existing image retrieval methods would suffice to verify the multimedia package integrity.

While information manipulation detection is a broad problem, in this paper we focus on verifying the \emph{semantic} integrity of multimedia packages. We define multimedia semantic integrity as the semantic consistency of information across all the modalities of a multimedia package.

We present a novel framework to solve a limited version of the multimedia information integrity assessment problem, where we consider each data package to contain only one image and an accompanying caption. Data packages in the reference dataset are used to train deep multimodal representation learning models (DMRLMs). The learned DMRLMs are then used to assess the integrity of query packages by calculating image-caption consistency scores (ICCSs) and employing outlier detection models (ODMs) to find their \textit{inlierness} with respect to the RD. We evaluate the proposed method on two publicly available datasets---Flickr30K~\cite{young_image_2014} and MS COCO~\cite{lin_microsoft_2014}, as well as on the MultimodAl Information Manipulation (MAIM) dataset that we created from image and caption pairs downloaded from Flickr, which we make publicly available.



To the best of our knowledge, ours is the first work to formally define the larger problem and provide an approach to solve it. Our work is significantly different from past work on robust hashing and watermarking~\cite{ababneh_scalable_2008,sun_secure_2014,wang_visual_2015,yan_multi-scale_2016} as those methods focus on the prevention of information manipulation while ours focuses on detection at a semantic level. The remainder of this paper is organized as follows. Section \ref{sec.related_work} reviews related work. In Section \ref{sec.seminteg} we describe the proposed method for assessing the semantic integrity of multimedia packages. In Section \ref{sec.data} we discuss existing public datasets as well as the new MAIM dataset. Experimental results and analysis are presented in Section \ref{sec.analysis}. Finally, in Section \ref{sec.conclusions}     we conclude the paper and introduce directions for future research. 

\section{Related Work}\label{sec.related_work}
Content integrity of multimedia data has been tackled in the past from the perspective of manipulation prevention using watermarking, authentication and hashing~\cite{ababneh_scalable_2008,sun_secure_2014,wang_visual_2015,yan_multi-scale_2016}. Most of such work is aimed at detecting tampered data, especially images, and approaches to recover the original data. This group of work focuses on the integrity of data with one modality. Our work is different in that it focuses on assessing the integrity across modalities of a multimodal package at a semantic level. The larger problem that we have defined above assumes that images are not tampered with but might be repurposed, thus creating fake data packages with inconsistent information.

Our methods in this paper are based on recent work in deep multimodal represenation learning~\cite{ngiam_multimodal_2011,vukotic_bidirectional_2016} and semantic retrieval of images and captions involving image-caption ranking~\cite{kiros_unifying_2014,frome_devise:_2013,weston_large_2010,hodosh_framing_2013,lin_leveraging_2016}. Deep representation learning performs very well at learning highly non-linear latent representations of data when large volumes of data are available. Autoencoders~\cite{hinton_reducing_2006} are a popular framework for unsupervised representation learning. Ngiam et al.~\cite{ngiam_multimodal_2011} showed how MAEs can be used to learn joint representations of data with multiple modalities. Vukoti\'c et al.~\cite{vukotic_bidirectional_2016} developed the BiDNN model that learns cross-modal mappings and joint representations of multimodal data.

Semantic retrieval of images from captions and vice versa has gained traction in recent years. Several methods have been developed that map images and captions to a common feature space so that their similarity, such as the cosine similarity, can be used to rank the affinity of image-caption pairs and return the top-\textit{K} candidates~\cite{kiros_unifying_2014,frome_devise:_2013,weston_large_2010,hodosh_framing_2013,lin_leveraging_2016}. Hodosh et al.~\cite{hodosh_framing_2013} use Kernel Canonical Correlation Analysis~\cite{bach_kernel_2002} to map image features and caption features to a common induced space. Wetson et al.~\cite{weston_large_2010} provide a method to simultaneously learn linear mappings from image and caption features to a common space with the objective of learning to associate images with correct captions. Frome et al.~\cite{frome_devise:_2013} present a deep visual-semantic embedding (DeViSE) model that learns to map image features to the space of caption features by optimizing a loss function that maximizes the cosine similarity of image-caption pairs while minimizing that of images and randomly picked text. The neural language model of Kiros et al.~\cite{kiros_unifying_2014} is based on a similar framework but learns to map caption features to the space of image features instead. In experiments, they showed that their model's performance is much better at the task of image-caption ranking compared to DeViSE.

Kiros et al.~\cite{kiros_unifying_2014} also compared their model to the multimodal recurrent neural network model of Mao et al.~\cite{mao_explain_2014} that automatically generates captions for images. This class of methods, based on caption generation, does not explicitly give a score for image-caption affinity and relies on perplexity when used for image-caption ranking or retrieval.

While previous work offers a way to rank image-caption pairs based on a measure of similarity, it does not provide a way to check the consistency of information between images and associated captions with respect to a reference dataset. Our work in this paper provides this novel contribution towards the larger goal of assessing the integrity of multimodal data packages.

\section{Semantic Integrity Assessment}\label{sec.seminteg}

\begin{figure*}
\centering
\begin{minipage}{0.45\textwidth}
\centering
\includegraphics[width=\textwidth,height=2in]{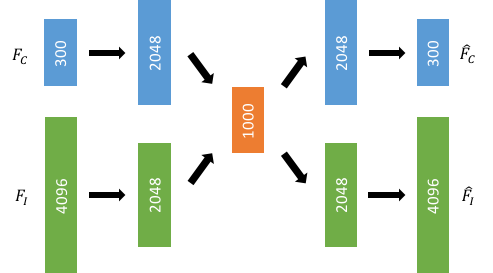}
\captionof{figure}{Our Multimodal Autoencoder Architecture. $F_I$ and $F_C$ are image and caption features respectively, while $\widehat{F}_I$ and $\widehat{F}_C$ are their reconstructed versions.}
\label{fig:mae}
\end{minipage} \hspace{0.5cm} %
\begin{minipage}{0.45\textwidth}
\centering
\includegraphics[width=\textwidth,height=2in]{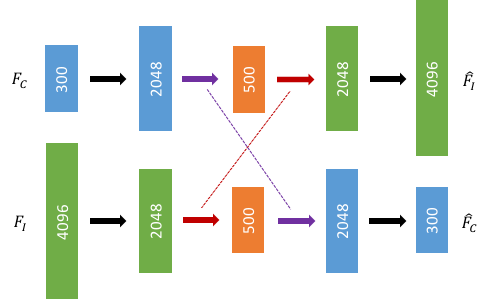}
\captionof{figure}{Our BiDNN Architecture. $F_I$ and $F_C$ are image and caption features respectively, while $\widehat{F}_I$ and $\widehat{F}_C$ are their reconstructed versions. Colored arrows with dotted connections reflect weight tying.}
\label{fig:bidnn}
\end{minipage}
\end{figure*}

One approach to information integrity assessment of a data object is to compare it against an existing knowledge-base (KB), with the assumption that such a KB exists. This KB can be explicit (such as a database of facts) or implicit (such as a learned statistical inference model). We use the observation that human beings verify the information integrity of pieces of data using world knowledge learned over time and external sources, such as an encyclopedia, to develop machine learning models that mimic world knowledge, and then use these models to assess the integrity of query data packages.

In order to verify the integrity of a query multimedia package that contains an image and an associated caption, we assume the existence of a reference set of similar media packages. This set, which we call the reference dataset (RD), serves as the KB to compare query packages against to measure their integrity. More specifically, we train an outlier detection model (ODM) on image-caption consistency scores (ICCSs) from packages in RD and use it to calculate the \textit{inlierness} of query packages. We employ deep multimodal representation learning models (DMRLMs) for jointly encoding images and corresponding captions, inspired by their success as reflected in recent literature, and use them to calculate ICCSs (depending on the DMRLM used). Fig.~\ref{fig:system} explains our complete integrity assessment system.

In this work we use a multimodal autoencoder (MAE)~\cite{ngiam_multimodal_2011}, a bidirectional (symmetrical) deep neural network (BiDNN)~\cite{vukotic_bidirectional_2016} or the unified visual semantic neural language model (VSM)~\cite{kiros_unifying_2014} as the embedding model. VGG19~\cite{simonyan_very_2014} image features are given as inputs to all these models, along with either average word2vec~\cite{mikolov_distributed_2013} embeddings (MAE and BiDNN) or one-hot encodings of words in captions (VSM).  The ODMs that we work with are the one-class support vector machine (OCSVM)~\cite{scholkopf_support_1999} and isolation forest (iForest)~\cite{liu_isolation_2008}. We discuss the aforementioned DMRLMs, with their associated ICCSs, and ODMs in detail in the following subsections.

\begin{figure*}
\centering
\includegraphics[width=0.8\textwidth]{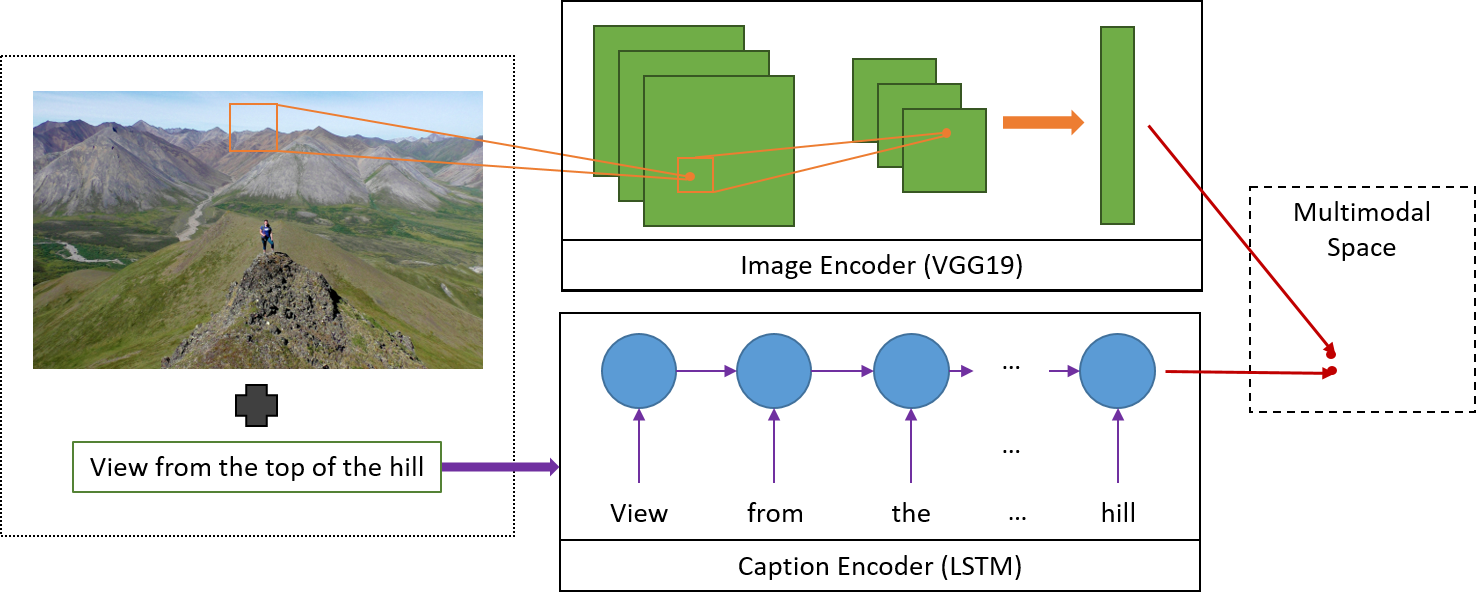}
\caption{The VSM Architecture of Kiros et al.~\cite{kiros_unifying_2014}}
\label{fig:uvsnlm}
\end{figure*}

\subsection{Deep Multimodal Representation Learning}

\subsubsection{Multimodal Autoencoder}

An autoencoder is a neural network that learns to reconstruct its input~\cite{hinton_reducing_2006}. Autoencoders are typically used to learn low-dimensional representations of data. The network architecture is designed such that the input goes through a series of layers with decreasing dimensionality to produce an encoding, which is then transformed through layers of increasing dimensionality to finally reconstruct the input. Ngiam et al.~\cite{ngiam_multimodal_2011} showed how an autoencoder network can be used to learn representations over multiple modalities. We train an MAE on the image-caption pairs in RD to learn their shared representation. Fig.~\ref{fig:mae} shows our MAE architecture, inspired by the bimodal deep autoencoder of Ngiam et al.~\cite{ngiam_multimodal_2011}. The image and caption features are passed through a series of unimodal layers before combining them in the shared representation layer. The decoder module of the MAE is a mirror image of its encoder. For MAE, we use reconstruction loss as the ICCS.


\begin{figure*}
\centering
\includegraphics[width=0.95\textwidth]{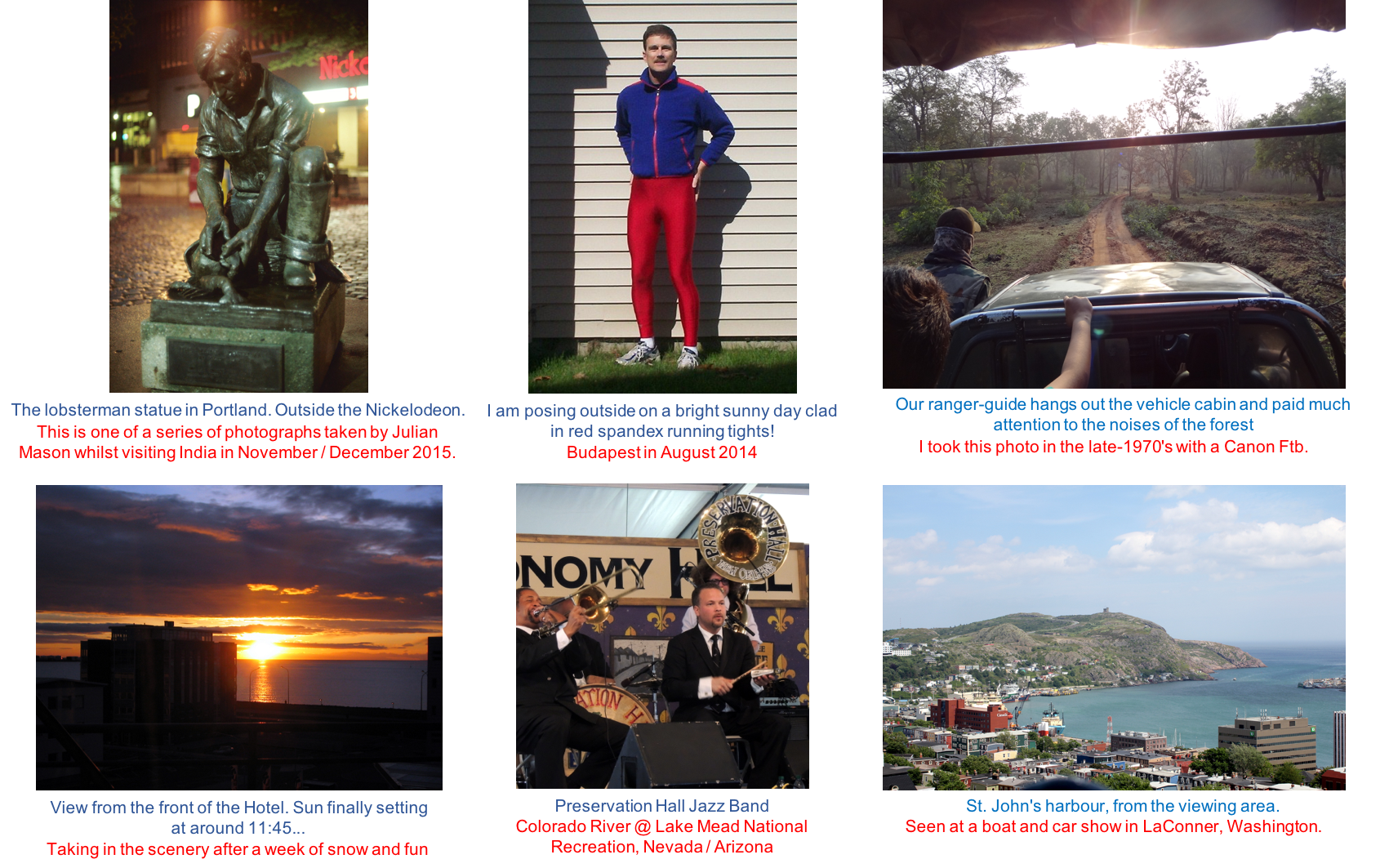}
\caption{Image-Caption Data package examples from MAIM dataset. The blue captions are the original ones that came with the image while the red ones are their manipulated versions.}
\label{fig:maim}
\end{figure*}

\subsubsection{Bidirectional (Symmetrical) Deep Neural Network}

A BiDNN is composed of two unimodal autoencoders with tied weights for the middle representation layers~\cite{vukotic_bidirectional_2016}. The network is trained to simultaneously reconstruct each modality from the other, learning cross-modal mappings as well as a joint representation. Fig.~\ref{fig:bidnn} shows our BiDNN architecture and illustrates the tied weights for a better understanding.  Our formulation of the joint representation is the same as Vukoti\'c et al.~\cite{vukotic_bidirectional_2016}, i.e., the concatenation of the activations of the two representation layers. We used the BiDNN package made available by Vukoti\'c et al.~\cite{vukotic_bidirectional_2016}\footnote{\url{https://github.com/v-v/BiDNN}} to implement our model. Reconstruction loss also serves as the ICCS in the case of BiDNN.

\subsubsection{Unified Visual Semantic Neural Language Model}

Kiros et al.~\cite{kiros_unifying_2014} introduced the unified visual semantic neural language model (VSM) that learns representations of captions in the embedding space of images, where image embeddings are first calculated using a deep neural network such as VGG19~\cite{simonyan_very_2014}. The VSM is trained to optimize a contrastive loss, which aims to maximize the cosine similarity between the representation of an image and the learned encoding of its caption while minimizing that between the image and captions not related to it. Fig.~\ref{fig:uvsnlm} shows the structure of the VSM. The network uses long short-term memory (LSTM) units~\cite{hochreiter_long_1997} to encode variable-length captions. We used the VSM package made available by Kiros et al.~\cite{kiros_unifying_2014}~\footnote{\url{https://github.com/ryankiros/visual-semantic-embedding}} and trained one model on each RD. Cosine similarity becomes the natural choice of ICCS in the case of VSM.

\subsection{Outlier Detection}

\subsubsection{One-Class Support Vector Machine}

The OCSVM is an unsupervised outlier detection model trained only on positive examples~\cite{scholkopf_support_1999}. It learns a decision function based on the distribution of the training data in its original or kernel space to classify the complete training dataset as the positive class, and everything else in the high-dimensional space as the negative class. This model is then used to predict whether a new data point is an inlier or an outlier with respect to the training data. This formulation of OCSVM fits very well with our approach of assessing the semantic integrity of a data package with respect to an RD (by using an OCSVM trained on the RD).

\makegapedcells
\begin{table*}
\caption{Evaluation Results on Flickr30K}
\label{tab:f30k}
\begin{tabular}{|c*{7} { | p{2cm} } |}
  \cline{3-8}
  \nocell{2}& \multicolumn{6}{c|}{Deep Multimodal Representation Learning Model} \\ \cline{3-8}
  \nocell{2} & \multicolumn{2}{c|}{MAE} & \multicolumn{2}{c|}{BiDNN} & \multicolumn{2}{c|}{VSM} \\ \cline{3-8}
  \nocell{2} & $F_1$-tampered & $F_1$-clean & $F_1$-tampered & $F_1$-clean & $F_1$-tampered & $F_1$-clean\\ \hline
  \multirowcell{2}[-0.75ex]{\rotatebox[origin=c]{90}{ODM}} & One-class SVM & 0.48 & 0.50 & 0.47 & 0.67 & \textbf{0.89} & \textbf{0.88} \\ \cline{2-8}
 & Isolation Forest & 0.49 & 0.50 & 0.63 & 0.62 & \textbf{0.81} & \textbf{0.7} \\ \cline{1-8}
\end{tabular}
\end{table*}

\makegapedcells
\begin{table*}
\caption{Evaluation Results on MS COCO}
\label{tab:mscoco}
\begin{tabular}{|c*{7} { | p{2cm} } |}
  \cline{3-8}
  \nocell{2}& \multicolumn{6}{c|}{Deep Multimodal Representation Learning Model} \\ \cline{3-8}
  \nocell{2} & \multicolumn{2}{c|}{MAE} & \multicolumn{2}{c|}{BiDNN} & \multicolumn{2}{c|}{VSM} \\ \cline{3-8}
  \nocell{2} & $F_1$-tampered & $F_1$-clean & $F_1$-tampered & $F_1$-clean & $F_1$-tampered & $F_1$-clean\\ \hline
  \multirowcell{2}[-0.75ex]{\rotatebox[origin=c]{90}{ODM}} & One-class SVM & 0.53 & 0.46 & 0.68 & 0.55 & \textbf{0.94} & \textbf{0.94} \\ \cline{2-8}
 & Isolation Forest & 0.5 & 0.48 & 0.76 & 0.77 & \textbf{0.94} & \textbf{0.94} \\ \cline{1-8}
\end{tabular}
\end{table*}

\makegapedcells
\begin{table*}
\caption{Evaluation Results on MAIM}
\label{tab:ours}
\begin{tabular}{|c*{7} { | p{2cm} } |}
  \cline{3-8}
  \nocell{2}& \multicolumn{6}{c|}{Deep Multimodal Representation Learning Model} \\ \cline{3-8}
  \nocell{2} & \multicolumn{2}{c|}{MAE} & \multicolumn{2}{c|}{BiDNN} & \multicolumn{2}{c|}{VSM} \\ \cline{3-8}
  \nocell{2} & $F_1$-tampered & $F_1$-clean & $F_1$-tampered & $F_1$-clean & $F_1$-tampered & $F_1$-clean\\ \hline
  \multirowcell{2}[-0.75ex]{\rotatebox[origin=c]{90}{ODM}} & One-class SVM & 0.49 & 0.49 & 0.46 & 0.5 & \textbf{0.75} & \textbf{0.76} \\ \cline{2-8}
 & Isolation Forest & 0.56 & 0.42 & 0.52 & 0.52 & \textbf{0.75} & \textbf{0.77} \\ \cline{1-8}
\end{tabular}
\end{table*}

\subsubsection{Isolation Forest}

An isolation forest (iForest) is a collection of decision trees that isolate a data point through recursive partitioning of random subsets of its features~\cite{liu_isolation_2008}. It works by first randomly selecting a feature of a data point and then finding a random split-value between its minimum and maximum values. This is then repeated recursively on the new splits. The recursive partitioning of a tree stops when a node contains only the provided data point. Under this setting, the average number of splits required (across all trees in the forest) to isolate a point gives an indication of its outlierness. The smaller the number, the higher the confidence that the point is an outlier; it is easier to isolate outliers as they lie in relatively low-density regions with respect to inliers (RD).

\section{Data}\label{sec.data}
We provide a quantitative evaluation of the performance of our method on three datasets: Flickr30K~\cite{young_image_2014}, MS COCO~\cite{lin_microsoft_2014} and a dataset that we created from images, captions and other metadata downloaded from Flickr (MAIM). While Flickr30K and MS COCO datasets contain objective captions which describe the contents of images, MAIM contains subjective captions, which do not necessarily do so and sometimes contain related information that might not be obvious from the image. Fig.~\ref{fig:maim} shows some examples from the MAIM dataset.

We use the training, validation and testing subsets of Flickr30K and MS COCO as made available by Kiros et al.~\cite{kiros_unifying_2014}\footnote{\url{https://github.com/ryankiros/visual-semantic-embedding}}, which makes sure that there is no overlap of images among the subsets. This is necessary because each image in these datasets has five captions (giving five image-caption pairs). There are $158,915$ and $423,915$ image-caption pairs in Flickr30K and MS COCO respectively, in total. Our dataset (MAIM) has $239,968$ image-caption pairs with exactly one caption for each unique image. We randomly split MAIM into training, validation and testing subsets and treat the training subset of each dataset as the RD in our framework. We replace the captions of half of the validation and testing images with captions of other images to create manipulated image-caption pairs for evaluation.

MAIM also has metadata for each package but we do not use them in our experiments in this work. This metadata includes location where the image was taken, time and date when the image was taken and information associated with the device used to capture the image.

\section{Analysis}\label{sec.analysis}
The inlier/outlier decisions of the ODMs in our system serve as the prediction of semantic information manipulation in query packages. We use $F_1$ scores as our evaluation metrics. Tables~\ref{tab:f30k},~\ref{tab:mscoco} and~\ref{tab:ours} summarize the results of our experiments on all combinations of DMRLMs and ODMs that we use in this work, on Flickr30K, MS COCO and MAIM respectively. We treat tampered packages as the positive class when calculating $F_1$-tampered and as the negative class for $F_1$-clean. The $F_1$-tampered and $F_1$-clean scores are coupled, i.e., every pair is from the same trained model.

We see that VSM consistently performs better than MAE and BiDNN in all our experiments on both metrics, with MAE consistently performing the worst. This gives us some key insights into the working of these DMRLMs. Even though MAE can compress multimodal data with low reconstruction error, it does not learn semantic associations between images and captions very well. The BiDNN model is trained to learn cross-modal mappings between images and captions, which forces it to learn those semantic associations. This explains why it works better than MAE at this task. The VSM model is trained to map captions to the representation space of images. The learning objective explicitly requires it to learn semantic relationships between the two modalities so that it can map captions consistent with an image close to it while inconsistent ones are mapped far from it. This makes VSM the strongest of the three models.

We also see that the $F_1$ scores of VSM on MS COCO are significantly better than those on the other datasets. This is expected and explained by the process through which the captions in the dataset were collected. Chen et al.~\cite{chen_microsoft_2015} used Amazon's Mechanical Turk\footnote{\url{https://www.mturk.com/mturk/welcome}} to gather objective captions with strong guidelines for their quality and content. The numbers are higher simply due to the better quality of captions and their objective content. This indicates that our method is better suited for objective captions.

\section{Conclusions and Future Work}\label{sec.conclusions}
Real-world multimedia data is often multimodal, consisting of images, videos, captions and other metadata. While multiple modalities present additional sources of information, it also makes such data packages vulnerable to tampering, where a subset of modalities might be manipulated, with possible malicious intent. In this paper we formally defined this problem and provided a method to solve a limited version of it (where each package has an image and a caption) as a first step towards the larger goal. Our method combines deep multimodal representation learning with outlier detection methods to assess whether a caption is consistent with the image in its package. We introduced the MultimodAl Information Manipulation dataset (MAIM) that we created for the larger problem, containing images, captions and various metadata, which we make available to the research community.\footnote{The code and data (MAIM) used in this work are available on request through email.} We presented a quantitative evaluation of our method on Flickr30K and MS COCO datasets, containing objective captions, and on the MAIM dataset, containing subjective captions. Our method was able to achieve $F_1$ scores of $0.75$, $0.89$ and $0.94$ on MAIM, Flickr30K and MS COCO, respectively, for detecting semantically incoherent media packages.

In our work we used the general formulation of MAE and BiDNN, providing these models VGG19 image features and aggregated word2vec caption features as inputs. It is possible that an end-to-end model with raw images and captions as inputs and a combination of convolution and recurrent layers might perform better. Similarly, training the image encoder of VSM jointly with the caption encoder might further boost its performance. We intend to explore these issues in future work. Our future work will also incorporate metadata and assess the integrity of entire packages. It is easy to see that our framework can be extended to include more modalities such as audio and video. We leave this to future work. This is in accordance with the larger goal of semantic multimodal information integrity assessment.

\begin{acks}

This work is based on research sponsored by the Defense Advanced Research Projects Agency under agreement number FA8750-16-2-0204. The U.S. Government is authorized to reproduce and distribute reprints for governmental purposes  notwithstanding any copyright notation thereon. The views and conclusions contained herein are those of the authors and should not be interpreted as necessarily representing the official policies or endorsements, either expressed or implied, of the Defense Advanced Research Projects Agency or the U.S. Government.

\end{acks}

\bibliographystyle{ACM-Reference-Format}
\bibliography{references} 


\begin{thebibliography}{00}


\ifx \showCODEN    \undefined \def \showCODEN     #1{\unskip}     \fi
\ifx \showDOI      \undefined \def \showDOI       #1{#1}\fi
\ifx \showISBNx    \undefined \def \showISBNx     #1{\unskip}     \fi
\ifx \showISBNxiii \undefined \def \showISBNxiii  #1{\unskip}     \fi
\ifx \showISSN     \undefined \def \showISSN      #1{\unskip}     \fi
\ifx \showLCCN     \undefined \def \showLCCN      #1{\unskip}     \fi
\ifx \shownote     \undefined \def \shownote      #1{#1}          \fi
\ifx \showarticletitle \undefined \def \showarticletitle #1{#1}   \fi
\ifx \showURL      \undefined \def \showURL       {\relax}        \fi
\providecommand\bibfield[2]{#2}
\providecommand\bibinfo[2]{#2}
\providecommand\natexlab[1]{#1}
\providecommand\showeprint[2][]{arXiv:#2}

\bibitem[\protect\citeauthoryear{Ababneh, Ansari, and Khokhar}{Ababneh
  et~al\mbox{.}}{2008}]%
        {ababneh_scalable_2008}
\bibfield{author}{\bibinfo{person}{Sufyan Ababneh}, \bibinfo{person}{Rashid
  Ansari}, {and} \bibinfo{person}{Ashfaq Khokhar}.}
  \bibinfo{year}{2008}\natexlab{}.
\newblock \showarticletitle{Scalable multimedia-content integrity verification
  with robust hashing}. In \bibinfo{booktitle}{{\em Electro/{Information}
  {Technology}, 2008. {EIT} 2008. {IEEE} {International} {Conference} on}}.
  \bibinfo{publisher}{IEEE}, \bibinfo{pages}{263--266}.
\newblock
\showURL{%
\url{http://ieeexplore.ieee.org/abstract/document/4554310/}}


\bibitem[\protect\citeauthoryear{Bach and Jordan}{Bach and Jordan}{2002}]%
        {bach_kernel_2002}
\bibfield{author}{\bibinfo{person}{Francis~R. Bach} {and}
  \bibinfo{person}{Michael~I. Jordan}.} \bibinfo{year}{2002}\natexlab{}.
\newblock \showarticletitle{Kernel independent component analysis}.
\newblock \bibinfo{journal}{{\em Journal of machine learning research\/}}
  \bibinfo{volume}{3}, \bibinfo{number}{Jul} (\bibinfo{year}{2002}),
  \bibinfo{pages}{1--48}.
\newblock
\showURL{%
\url{http://www.jmlr.org/papers/v3/bach02a}}


\bibitem[\protect\citeauthoryear{Chen, Fang, Lin, Vedantam, Gupta, Dollár, and
  Zitnick}{Chen et~al\mbox{.}}{2015}]%
        {chen_microsoft_2015}
\bibfield{author}{\bibinfo{person}{Xinlei Chen}, \bibinfo{person}{Hao Fang},
  \bibinfo{person}{Tsung-Yi Lin}, \bibinfo{person}{Ramakrishna Vedantam},
  \bibinfo{person}{Saurabh Gupta}, \bibinfo{person}{Piotr Dollár}, {and}
  \bibinfo{person}{C.~Lawrence Zitnick}.} \bibinfo{year}{2015}\natexlab{}.
\newblock \showarticletitle{Microsoft {COCO} captions: {Data} collection and
  evaluation server}.
\newblock \bibinfo{journal}{{\em arXiv preprint arXiv:1504.00325\/}}
  (\bibinfo{year}{2015}).
\newblock
\showURL{%
\url{https://arxiv.org/abs/1504.00325}}


\bibitem[\protect\citeauthoryear{Frome, Corrado, Shlens, Bengio, Dean, Mikolov,
  and {others}}{Frome et~al\mbox{.}}{2013}]%
        {frome_devise:_2013}
\bibfield{author}{\bibinfo{person}{Andrea Frome}, \bibinfo{person}{Greg~S.
  Corrado}, \bibinfo{person}{Jon Shlens}, \bibinfo{person}{Samy Bengio},
  \bibinfo{person}{Jeff Dean}, \bibinfo{person}{Tomas Mikolov}, {and}
  \bibinfo{person}{{others}}.} \bibinfo{year}{2013}\natexlab{}.
\newblock \showarticletitle{Devise: {A} deep visual-semantic embedding model}.
  In \bibinfo{booktitle}{{\em Advances in neural information processing
  systems}}. \bibinfo{pages}{2121--2129}.
\newblock
\showURL{%
\url{http://papers.nips.cc/paper/5204-devise-a-deep-visual-semantic-embedding-model}}


\bibitem[\protect\citeauthoryear{Hinton and Salakhutdinov}{Hinton and
  Salakhutdinov}{2006}]%
        {hinton_reducing_2006}
\bibfield{author}{\bibinfo{person}{Geoffrey~E. Hinton} {and}
  \bibinfo{person}{Ruslan~R. Salakhutdinov}.} \bibinfo{year}{2006}\natexlab{}.
\newblock \showarticletitle{Reducing the dimensionality of data with neural
  networks}.
\newblock \bibinfo{journal}{{\em science\/}} \bibinfo{volume}{313},
  \bibinfo{number}{5786} (\bibinfo{year}{2006}), \bibinfo{pages}{504--507}.
\newblock
\showURL{%
\url{http://science.sciencemag.org/content/313/5786/504.short}}


\bibitem[\protect\citeauthoryear{Hochreiter and Schmidhuber}{Hochreiter and
  Schmidhuber}{1997}]%
        {hochreiter_long_1997}
\bibfield{author}{\bibinfo{person}{Sepp Hochreiter} {and}
  \bibinfo{person}{Jürgen Schmidhuber}.} \bibinfo{year}{1997}\natexlab{}.
\newblock \showarticletitle{Long short-term memory}.
\newblock \bibinfo{journal}{{\em Neural computation\/}} \bibinfo{volume}{9},
  \bibinfo{number}{8} (\bibinfo{year}{1997}), \bibinfo{pages}{1735--1780}.
\newblock
\showURL{%
\url{http://www.mitpressjournals.org/doi/abs/10.1162/neco.1997.9.8.1735}}


\bibitem[\protect\citeauthoryear{Hodosh, Young, and Hockenmaier}{Hodosh
  et~al\mbox{.}}{2013}]%
        {hodosh_framing_2013}
\bibfield{author}{\bibinfo{person}{Micah Hodosh}, \bibinfo{person}{Peter
  Young}, {and} \bibinfo{person}{Julia Hockenmaier}.}
  \bibinfo{year}{2013}\natexlab{}.
\newblock \showarticletitle{Framing image description as a ranking task:
  {Data}, models and evaluation metrics}.
\newblock \bibinfo{journal}{{\em Journal of Artificial Intelligence
  Research\/}}  \bibinfo{volume}{47} (\bibinfo{year}{2013}),
  \bibinfo{pages}{853--899}.
\newblock
\showURL{%
\url{http://www.jair.org/papers/paper3994.html}}


\bibitem[\protect\citeauthoryear{Kiros, Salakhutdinov, and Zemel}{Kiros
  et~al\mbox{.}}{2014}]%
        {kiros_unifying_2014}
\bibfield{author}{\bibinfo{person}{Ryan Kiros}, \bibinfo{person}{Ruslan
  Salakhutdinov}, {and} \bibinfo{person}{Richard~S. Zemel}.}
  \bibinfo{year}{2014}\natexlab{}.
\newblock \showarticletitle{Unifying visual-semantic embeddings with multimodal
  neural language models}.
\newblock \bibinfo{journal}{{\em arXiv preprint arXiv:1411.2539\/}}
  (\bibinfo{year}{2014}).
\newblock
\showURL{%
\url{https://arxiv.org/abs/1411.2539}}


\bibitem[\protect\citeauthoryear{Lin, Maire, Belongie, Hays, Perona, Ramanan,
  Dollár, and Zitnick}{Lin et~al\mbox{.}}{2014}]%
        {lin_microsoft_2014}
\bibfield{author}{\bibinfo{person}{Tsung-Yi Lin}, \bibinfo{person}{Michael
  Maire}, \bibinfo{person}{Serge Belongie}, \bibinfo{person}{James Hays},
  \bibinfo{person}{Pietro Perona}, \bibinfo{person}{Deva Ramanan},
  \bibinfo{person}{Piotr Dollár}, {and} \bibinfo{person}{C.~Lawrence
  Zitnick}.} \bibinfo{year}{2014}\natexlab{}.
\newblock \showarticletitle{Microsoft coco: {Common} objects in context}. In
  \bibinfo{booktitle}{{\em European {Conference} on {Computer} {Vision}}}.
  \bibinfo{publisher}{Springer}, \bibinfo{pages}{740--755}.
\newblock
\showURL{%
\url{http://link.springer.com/chapter/10.1007/978-3-319-10602-1_48}}


\bibitem[\protect\citeauthoryear{Lin and Parikh}{Lin and Parikh}{2016}]%
        {lin_leveraging_2016}
\bibfield{author}{\bibinfo{person}{Xiao Lin} {and} \bibinfo{person}{Devi
  Parikh}.} \bibinfo{year}{2016}\natexlab{}.
\newblock \showarticletitle{Leveraging visual question answering for
  image-caption ranking}. In \bibinfo{booktitle}{{\em European {Conference} on
  {Computer} {Vision}}}. \bibinfo{publisher}{Springer},
  \bibinfo{pages}{261--277}.
\newblock
\showURL{%
\url{http://link.springer.com/chapter/10.1007/978-3-319-46475-6_17}}


\bibitem[\protect\citeauthoryear{Liu, Ting, and Zhou}{Liu
  et~al\mbox{.}}{2008}]%
        {liu_isolation_2008}
\bibfield{author}{\bibinfo{person}{Fei~Tony Liu}, \bibinfo{person}{Kai~Ming
  Ting}, {and} \bibinfo{person}{Zhi-Hua Zhou}.}
  \bibinfo{year}{2008}\natexlab{}.
\newblock \showarticletitle{Isolation forest}. In \bibinfo{booktitle}{{\em Data
  {Mining}, 2008. {ICDM}'08. {Eighth} {IEEE} {International} {Conference} on}}.
  \bibinfo{publisher}{IEEE}, \bibinfo{pages}{413--422}.
\newblock
\showURL{%
\url{http://ieeexplore.ieee.org/abstract/document/4781136/}}


\bibitem[\protect\citeauthoryear{Mao, Xu, Yang, Wang, and Yuille}{Mao
  et~al\mbox{.}}{2014}]%
        {mao_explain_2014}
\bibfield{author}{\bibinfo{person}{Junhua Mao}, \bibinfo{person}{Wei Xu},
  \bibinfo{person}{Yi Yang}, \bibinfo{person}{Jiang Wang}, {and}
  \bibinfo{person}{Alan~L. Yuille}.} \bibinfo{year}{2014}\natexlab{}.
\newblock \showarticletitle{Explain images with multimodal recurrent neural
  networks}.
\newblock \bibinfo{journal}{{\em arXiv preprint arXiv:1410.1090\/}}
  (\bibinfo{year}{2014}).
\newblock
\showURL{%
\url{https://arxiv.org/abs/1410.1090}}


\bibitem[\protect\citeauthoryear{Mikolov, Sutskever, Chen, Corrado, and
  Dean}{Mikolov et~al\mbox{.}}{2013}]%
        {mikolov_distributed_2013}
\bibfield{author}{\bibinfo{person}{Tomas Mikolov}, \bibinfo{person}{Ilya
  Sutskever}, \bibinfo{person}{Kai Chen}, \bibinfo{person}{Greg~S. Corrado},
  {and} \bibinfo{person}{Jeff Dean}.} \bibinfo{year}{2013}\natexlab{}.
\newblock \showarticletitle{Distributed representations of words and phrases
  and their compositionality}. In \bibinfo{booktitle}{{\em Advances in neural
  information processing systems}}. \bibinfo{pages}{3111--3119}.
\newblock


\bibitem[\protect\citeauthoryear{Ngiam, Khosla, Kim, Nam, Lee, and Ng}{Ngiam
  et~al\mbox{.}}{2011}]%
        {ngiam_multimodal_2011}
\bibfield{author}{\bibinfo{person}{Jiquan Ngiam}, \bibinfo{person}{Aditya
  Khosla}, \bibinfo{person}{Mingyu Kim}, \bibinfo{person}{Juhan Nam},
  \bibinfo{person}{Honglak Lee}, {and} \bibinfo{person}{Andrew~Y. Ng}.}
  \bibinfo{year}{2011}\natexlab{}.
\newblock \showarticletitle{Multimodal deep learning}. In
  \bibinfo{booktitle}{{\em Proceedings of the 28th international conference on
  machine learning ({ICML}-11)}}. \bibinfo{pages}{689--696}.
\newblock
\showURL{%
\url{http://machinelearning.wustl.edu/mlpapers/paper_files/ICML2011Ngiam_399.pdf}}


\bibitem[\protect\citeauthoryear{Schölkopf, Williamson, Smola, Shawe-Taylor,
  Platt, and {others}}{Schölkopf et~al\mbox{.}}{1999}]%
        {scholkopf_support_1999}
\bibfield{author}{\bibinfo{person}{Bernhard Schölkopf},
  \bibinfo{person}{Robert~C. Williamson}, \bibinfo{person}{Alexander~J. Smola},
  \bibinfo{person}{John Shawe-Taylor}, \bibinfo{person}{John~C. Platt}, {and}
  \bibinfo{person}{{others}}.} \bibinfo{year}{1999}\natexlab{}.
\newblock \showarticletitle{Support vector method for novelty detection.}. In
  \bibinfo{booktitle}{{\em {NIPS}}}, Vol.~\bibinfo{volume}{12}.
  \bibinfo{pages}{582--588}.
\newblock
\showURL{%
\url{https://papers.nips.cc/paper/1723-support-vector-method-for-novelty-detection.pdf}}


\bibitem[\protect\citeauthoryear{Simonyan and Zisserman}{Simonyan and
  Zisserman}{2014}]%
        {simonyan_very_2014}
\bibfield{author}{\bibinfo{person}{Karen Simonyan} {and}
  \bibinfo{person}{Andrew Zisserman}.} \bibinfo{year}{2014}\natexlab{}.
\newblock \showarticletitle{Very deep convolutional networks for large-scale
  image recognition}.
\newblock \bibinfo{journal}{{\em arXiv preprint arXiv:1409.1556\/}}
  (\bibinfo{year}{2014}).
\newblock
\showURL{%
\url{https://arxiv.org/abs/1409.1556}}


\bibitem[\protect\citeauthoryear{Sun and Zeng}{Sun and Zeng}{2014}]%
        {sun_secure_2014}
\bibfield{author}{\bibinfo{person}{Rui Sun} {and} \bibinfo{person}{Wenjun
  Zeng}.} \bibinfo{year}{2014}\natexlab{}.
\newblock \showarticletitle{Secure and robust image hashing via compressive
  sensing}.
\newblock \bibinfo{journal}{{\em Multimedia tools and applications\/}}
  \bibinfo{volume}{70}, \bibinfo{number}{3} (\bibinfo{year}{2014}),
  \bibinfo{pages}{1651--1665}.
\newblock
\showURL{%
\url{http://link.springer.com/article/10.1007/s11042-012-1188-8}}


\bibitem[\protect\citeauthoryear{Vukotić, Raymond, and Gravier}{Vukotić
  et~al\mbox{.}}{2016}]%
        {vukotic_bidirectional_2016}
\bibfield{author}{\bibinfo{person}{Vedran Vukotić}, \bibinfo{person}{Christian
  Raymond}, {and} \bibinfo{person}{Guillaume Gravier}.}
  \bibinfo{year}{2016}\natexlab{}.
\newblock \showarticletitle{Bidirectional joint representation learning with
  symmetrical deep neural networks for multimodal and crossmodal applications}.
  In \bibinfo{booktitle}{{\em Proceedings of the 2016 {ACM} on {International}
  {Conference} on {Multimedia} {Retrieval}}}. \bibinfo{publisher}{ACM},
  \bibinfo{pages}{343--346}.
\newblock
\showURL{%
\url{http://dl.acm.org/citation.cfm?id=2912064}}


\bibitem[\protect\citeauthoryear{Wang, Pang, Zhou, Zhou, Li, and Xue}{Wang
  et~al\mbox{.}}{2015}]%
        {wang_visual_2015}
\bibfield{author}{\bibinfo{person}{Xiaofeng Wang}, \bibinfo{person}{Kemu Pang},
  \bibinfo{person}{Xiaorui Zhou}, \bibinfo{person}{Yang Zhou},
  \bibinfo{person}{Lu Li}, {and} \bibinfo{person}{Jianru Xue}.}
  \bibinfo{year}{2015}\natexlab{}.
\newblock \showarticletitle{A visual model-based perceptual image hash for
  content authentication}.
\newblock \bibinfo{journal}{{\em IEEE Transactions on Information Forensics and
  Security\/}} \bibinfo{volume}{10}, \bibinfo{number}{7}
  (\bibinfo{year}{2015}), \bibinfo{pages}{1336--1349}.
\newblock
\showURL{%
\url{http://ieeexplore.ieee.org/abstract/document/7050251/}}


\bibitem[\protect\citeauthoryear{Weston, Bengio, and Usunier}{Weston
  et~al\mbox{.}}{2010}]%
        {weston_large_2010}
\bibfield{author}{\bibinfo{person}{Jason Weston}, \bibinfo{person}{Samy
  Bengio}, {and} \bibinfo{person}{Nicolas Usunier}.}
  \bibinfo{year}{2010}\natexlab{}.
\newblock \showarticletitle{Large scale image annotation: learning to rank with
  joint word-image embeddings}.
\newblock \bibinfo{journal}{{\em Machine learning\/}} \bibinfo{volume}{81},
  \bibinfo{number}{1} (\bibinfo{year}{2010}), \bibinfo{pages}{21--35}.
\newblock
\showURL{%
\url{http://www.springerlink.com/index/Y277128518468756.pdf}}


\bibitem[\protect\citeauthoryear{Yan, Pun, and Yuan}{Yan et~al\mbox{.}}{2016}]%
        {yan_multi-scale_2016}
\bibfield{author}{\bibinfo{person}{Cai-Ping Yan}, \bibinfo{person}{Chi-Man
  Pun}, {and} \bibinfo{person}{Xiao-Chen Yuan}.}
  \bibinfo{year}{2016}\natexlab{}.
\newblock \showarticletitle{Multi-scale image hashing using adaptive local
  feature extraction for robust tampering detection}.
\newblock \bibinfo{journal}{{\em Signal Processing\/}}  \bibinfo{volume}{121}
  (\bibinfo{year}{2016}), \bibinfo{pages}{1--16}.
\newblock
\showURL{%
\url{http://www.sciencedirect.com/science/article/pii/S0165168415003709}}


\bibitem[\protect\citeauthoryear{Young, Lai, Hodosh, and Hockenmaier}{Young
  et~al\mbox{.}}{2014}]%
        {young_image_2014}
\bibfield{author}{\bibinfo{person}{Peter Young}, \bibinfo{person}{Alice Lai},
  \bibinfo{person}{Micah Hodosh}, {and} \bibinfo{person}{Julia Hockenmaier}.}
  \bibinfo{year}{2014}\natexlab{}.
\newblock \showarticletitle{From image descriptions to visual denotations:
  {New} similarity metrics for semantic inference over event descriptions}.
\newblock \bibinfo{journal}{{\em Transactions of the Association for
  Computational Linguistics\/}}  \bibinfo{volume}{2} (\bibinfo{year}{2014}),
  \bibinfo{pages}{67--78}.
\newblock
\showURL{%
\url{https://www.transacl.org/ojs/index.php/tacl/article/view/229}}


\end{thebibliography}

\end{document}